\documentclass[11pt,a4paper]{article}
\usepackage{amsmath,amssymb}
\usepackage{epsfig,graphicx}
\usepackage{cite}
\usepackage{amsbsy}

\def\mprp{\mbox{\tiny $\bot$}}
\def\mprl{\mbox{\tiny $\|$}}

\newcommand{\bs}{\boldsymbol} 
\newcommand{\D}{\mathrm{d}} 

\begin{document}

\title{\bf Neutrino production of electron-positron pairs at~excited Landau levels in a strong magnetic field} 

\author{A.~V.~Kuznetsov$^a$\footnote{{\bf e-mail}: avkuzn@uniyar.ac.ru},
D.~A.~Rumyantsev$^a$\footnote{{\bf e-mail}: rda@uniyar.ac.ru},
V.~N.~Savin$^b$\footnote{{\bf e-mail}: vs\_post07@mail.ru}
\\
$^a$ \small{\em Yaroslavl State P.G.~Demidov University} \\
\small{\em Sovietskaya 14, 150000 Yaroslavl, Russian Federation }
\\
$^b$ \small{\em A.F.~Mozhaiskiy Space Military Academy, Yaroslavl Branch} \\
\small{\em Moskovskiy Prosp. 28, 150001 Yaroslavl, Russian Federation }
}

\date{}

\maketitle

\begin{abstract}
The process of neutrino production of electron positron pairs in a magnetic field of arbitrary strength, where 
electrons and positrons can be created in the states corresponding to excited Landau levels, is analysed. 
The mean value of the neutrino energy loss due to the process 
$\nu \to \nu e^- e^+$ is calculated. The result can be applied for calculating the efficiency of the electron-positron plasma production by neutrinos 
in the conditions of the Kerr black hole accretion disc considered by experts as the most possible source of a short 
cosmological gamma burst. The presented research can be also 
useful for further development of the calculation technic for an analysis of quantum processes 
in external active medium, and in part in the conditions of moderately strong magnetic field, when taking 
account of the ground Landau level appears to be insufficient.
\end{abstract}

\section{Introduction}
\label{sec:Introduction}

An intense electromagnetic field makes possible the processes which are forbidden in a vacuum such as 
the neutrino production of an electron--positron pair $\nu \to \nu e^- e^+$. 
The list of papers devoted to an analysis of this process and the collection of the results obtained 
could be found e.g. in Ref.~\cite{KM_Book_2013}. 
In most cases, calculations of this kind were made either in the crossed field approximation, 
or in the limit of a superstrong field much greater than the critical value of 
$B_e = m_e^2/e \simeq 4.41\times 10^{13}$~G, 
when the electrons and positrons were born
in states corresponding to the ground Landau level. 
However, there are physical situations of the so-called moderately strong magnetic 
field\footnote{We use natural units
$c = \hbar = k_{\rm{B}} = 1$, $m_e$ is the electron mass, and $e$ is the elementary
charge.}, 
$p_\perp^2 \gtrsim e B \gg m_e^2$, when electrons and positrons mainly occupy the ground
Landau level, however, a noticeable fraction may be produced at the next levels. 

The indicated hierarchy of physical parameters corresponds to the conditions of the Kerr black hole accretion disk, regarded by experts as the most likely source of a short cosmological gamma-ray burst. 
The disc is a source of copious neutrinos and anti-neutrinos, which partially annihilate above the disc
and turn into $e^{\mp}$ pairs, $\nu \bar\nu \to e^- e^+$. This process was proposed and investigated in many details, see e.g. Ref.~\cite{Birkl:2007} and the papers cited therein,
as a possible mechanism for creating relativistic, $e^{\mp}$-dominated jets that could power observed 
gamma-ray bursts. 
In Ref.~\cite{Beloborodov:2011}, in addition to $\nu \bar\nu$ annihilation, the contribution
of the magnetic field-induced process $\nu \to \nu e^- e^+$ to the neutrino energy deposition rate around
the black hole was also included. 

However, 
in calculations of the efficiency of the electron-positron plasma production by neutrino through 
the process $\nu \to \nu e^- e^+$ in those physical conditions~\cite{Beloborodov:2011} 
($B$ to 180 $B_e$, $E$ to 25 MeV), 
it should be kept in mind that approximations 
of both the crossed and superstrong field have a limited applicability here. 
We know a limited number of papers~\cite{Bezchastnov:1996,Mikheev:2000,Dicus:2007}, 
where the probability of neutrino-electron processes was investigated,
as the sum over the Landau levels of electrons (positrons).
In the papers~\cite{Bezchastnov:1996,Mikheev:2000},
only the neutrino-electron scattering channel in a dense magnetized plasma was studied, 
which was the crossed process to the considered here neutrino creation of electron-positron pairs.
In the paper~\cite{Dicus:2007}, also devoted to the study of the process $\nu \to \nu e^- e^+$, the analytical
calculations were presented in a rather cumbersome form, caused by the choice of solutions of the Dirac equation.
The final results for the process probability were obtained by numerical calculations for some set of Landau levels 
occupied by electrons and positrons. 
In astrophysical applications, there exists probably more interesting value than the process probability, 
namely, the mean value of the neutrino energy loss, caused by the influence of an external magnetic field. 

Thus, the aim of this paper is the study of the process $\nu \to \nu e^- e^+$ in the physical conditions of the moderately strong 
magnetic field, where the electrons and positrons would be born in the states corresponding to the excited
Landau levels, and the theoretical description would contain a relatively simple analytical formulas 
for the mean value of the neutrino energy loss, for a wide range of Landau levels. 
More details of the analysis can be found in our recent paper~\cite{Kuznetsov:2014}. 

\section{The probability of the process $\nu \to \nu e^- e^+$}	
\label{sec:probability}

We use the standard calculation technics, see e.g. Ref.~\cite{KM_Book_2013}. 
The effective local Lagrangian of the process can be written in the form
\begin{equation}
{\cal L} \, = \, - \frac{G_{\mathrm{F}}}{\sqrt 2} 
\big [ \bar e \gamma_{\alpha} (C_V - C_A \gamma_5) e \big ] \,
\big [ \bar \nu \gamma^{\alpha} (1 - \gamma_5) \nu \big ] \,,
\label{eq:L}
\end{equation}
where the electron field operators are constructed on a base of the solutions of the Dirac equation 
in the presence of an external magnetic field. 
The constants $C_V$ and $C_A$ for different neutrino types are:
\begin{eqnarray}
&&C_V^{(e)} = + \frac{1}{2} + 2 \sin^2 \theta_{\rm W} \,, \qquad C_A^{(e)} = + \frac{1}{2} \,,
\nonumber\\
&&C_V^{(\mu,\tau)} = - \frac{1}{2} + 2 \sin^2 \theta_{\rm W} \,, \qquad C_A^{(\mu,\tau)} = - \frac{1}{2} \,,
\label{eq:CVCA}
\end{eqnarray}
where $\theta_\mathrm{W}$ is the Weinberg angle.

The total probability of the process $\nu \to \nu e^-_{(n)} e^+_{(\ell)}$ where the
electron and the positron are created in the states corresponding to $n$th and $\ell$th Landau levels
correspondingly, is, in a general case, the sum of the probabilities of the four polarization channels:
\begin{equation}
\label{eq:Wtot}
W_{n \ell} = W^{--}_{n \ell} + W^{-+}_{n \ell} + W^{+-}_{n \ell} + W^{++}_{n \ell} \, .
\end{equation}
For each of the channels, the differential probability over the final neutrino momentum 
per unit time can be written as
\begin{equation}
\D W^{s s'}_{n \ell} \, = \, \frac{1}{\cal T}\, 
\frac{\D^3 P'\,V}{(2 \pi)^3} \; 
\int \, |{\cal S}_{n \ell}^{s s'}|^2 \, 
\D \Gamma_{e^-} \;
\D \Gamma_{e^+} \,,
\label{eq:dw1} 
\end{equation}
where $\cal T$ is the total interaction time, $V = L_x L_y L_z$ is the total volume 
of the interaction region, 
${\cal S}_{n \ell}^{s s'}$ is the $S$-matrix element constructed with the effective Lagrangian~(\ref{eq:L}),
and the elements of the phase volume are introduced for the electron and the positron 
(the magnetic field is directed along the $z$ axis):
\begin{equation}
\D \Gamma_{e^-} = \frac{\D^2 p\,L_y L_z}{(2 \pi)^2}, \quad
\D \Gamma_{e^+} = \frac{\D^2 p'\,L_y L_z}{(2 \pi)^2} \, .
\label{eq:d_n} 
\end{equation}

In the integration over the momenta of the electron and the positron, a condition arises: 
\begin{equation}
q_{\mprl}^2 = q_0^2 - q_z^2 \geqslant (M_{n} + M_{\ell})^2 \, ,
\label{eq:cond} 
\end{equation}
which determines the range of integration over the final neutrino momentum. 
Here, $q = P - P' = p + p'$ is the change of the four-vector of the neutrino 
momentum equal to the four-momentum of the $e^- e^+$ pair, $P^\alpha = (E, {\bs P})$ and 
$P\,'^\alpha = (E\,', {\bs P}\,')$ are the four-momenta of the initial and final neutrinos. 

In turn, the condition~(\ref{eq:cond}) can be satisfied when the energy of the initial
neutrino exceeds a certain threshold value.
In the reference frame, hereafter called $K$, where the momentum of the initial neutrino directed at an angle
$\theta$ to the magnetic field, the threshold energy is given by:
\begin{equation}
E \, \sin \theta \geqslant M_{n} + M_{\ell} \, .
\label{eq:condE} 
\end{equation}

\begin{figure}[!t] 
\centering
\begin{minipage}{7cm}
\includegraphics[scale=0.75]{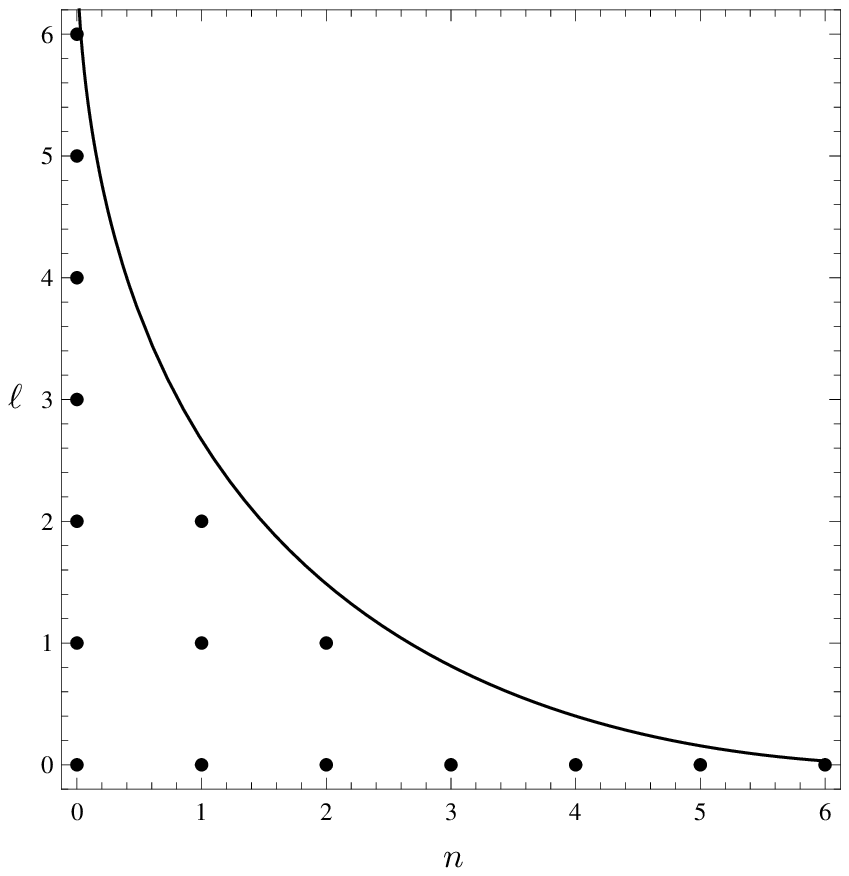}
\end{minipage}
\qquad
\begin{minipage}{7cm}
\includegraphics[scale=0.75]{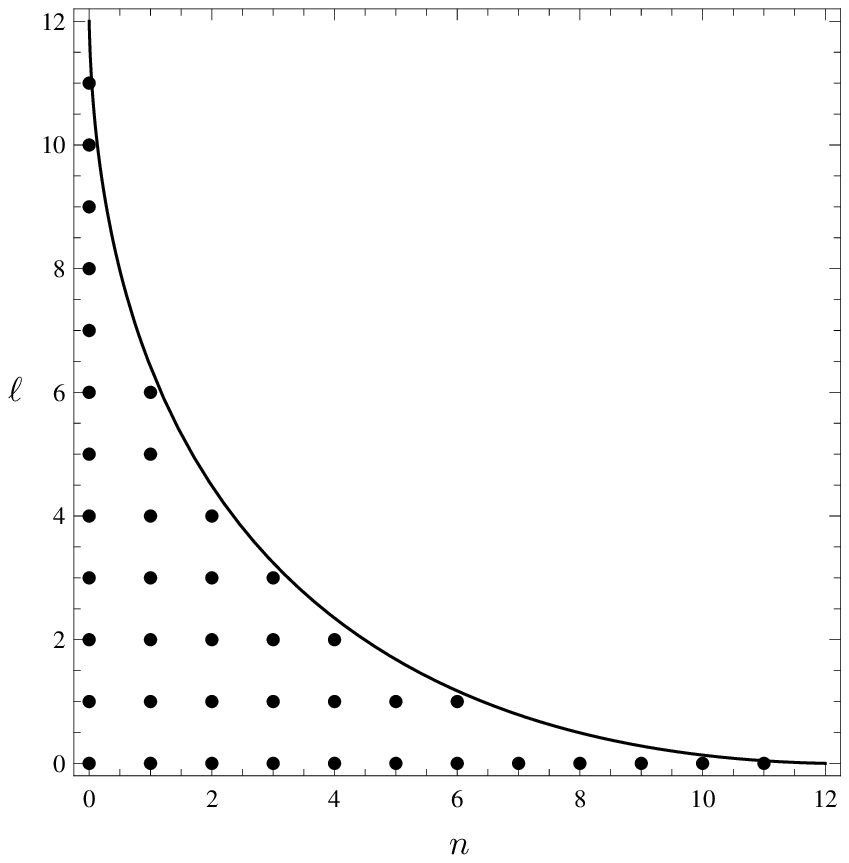}
\end{minipage}
\caption{Landau levels of $e^-_{(n)} e^+_{(\ell)}$ to be excited when $M_{n} + M_{\ell} \leqslant E_{\mprp}$,
at $E_{\mprp} = 25$ MeV, and at $B =  180 \,B_e$ (left) and $100 \,B_e$ (right).}
\label{fig:open_levels}
\end{figure}

In Fig.~\ref{fig:open_levels}, the Landau levels of $e^-_{(n)} e^+_{(\ell)}$ are shown, 
similarly to Fig.~1 of Ref.~\cite{Daugherty:1983}, to be excited 
accoring to the condition~(\ref{eq:condE}), 
at $E \, \sin \theta = 25$ MeV, and at $B =  180 \,B_e$ and $100 \,B_e$.

It is convenient to perform further integration over the final neutrino momentum, 
without loss of generality, not in an arbitrary reference frame $K$, 
but in the special frame $K_0$, 
where the initial neutrino momentum is perpendicular to the magnetic
field, $P_z = 0$. One can then return from 
$K_0$ to $K$ by the Lorentz transformation along the field
(recall that the field is invariant with respect to this transformation). 

It is convenient to use the dimensionless cylindrical
coordinates in the space of the final neutrino momentum vector $\bs P\,'$:
\begin{eqnarray}
&&\rho = \sqrt{P_x'^2 + P_y'^2}/E_{\mprp} \,, 
\qquad \tan \,\phi = P_y'/P_x'\,, 
\qquad z = P_z'/E_{\mprp} \,,
\nonumber\\[2mm]
&&r = E'/E_{\mprp} = \sqrt{\rho^2 + z^2} \,.
\label{eq:variab}
\end{eqnarray}
Here, $E_{\mprp}$ is the energy of the initial neutrino in the frame $K_0$, 
which is connected with its energy $E$ in an arbitrary frame $K$ by the 
relation $E_{\mprp} = E \sin \theta$.  

We do not present here the set of expressions for the probability of the process $\nu \to \nu e^-_{(n)} e^+_{(\ell)}$, 
which can be found in the paper~\cite{Kuznetsov:2014}.
These probabilities evaluated numerically as the functions of the initial neutrino energy and 
of the magnetic field strength for all channels considered in 
Ref.~\cite{Dicus:2007}, where the electron and positron are created 
in the lower Landau levels, are in a good agreement with the results of that paper.

\section{Neutrino energy and momentum losses}	
\label{sec:losses}

The probability of the $\nu \to \nu e^- e^+$ process defines 
its partial contribution into the neutrino opacity of the medium. 
The estimation of the neutrino mean free 
path with respect to this process gives the result which is too large~\cite{KM_Book_2013} 
compared with the typical size of a compact astrophysical object, e.g. the supernova remnant, 
where a strong magnetic field could exist. 
However, a mean free path does not exhaust the neutrino physics in 
a medium. In astrophysical applications, we could consider 
the values that probably are more essential, namely, the mean values 
of the neutrino energy and momentum losses, caused by the influence of an external magnetic field. 
These values can be described by the four-vector of losses $Q^{\alpha}$, 
\begin{equation}
Q^\alpha \, = \, E \int q^\alpha \, \D W = - E \, ({\cal I}, {\bs F}) \,.
\label{eq:Q0}
\end{equation}
where $\D W$ is the total differential probability of the process $\nu \to \nu e^- e^+$. 
The zeroth component of $Q^{\alpha}$ is connected with the mean energy lost 
by a neutrino per unit time due to the process considered, 
${\cal I} = \D E/\D t$.
The space components of the four-vector~(\ref{eq:Q0}) are similarly 
connected with the mean neutrino momentum loss per unit time, 
${\bs F} = \D {\bs P}/\D t$. 
It should be noted that the four-vector of losses $Q^{\alpha}$ can be used for evaluating 
the integral effect of neutrinos on plasma 
in the conditions of not very dense plasma, e.g. of a supernova envelope, 
when an one-interaction approximation of a neutrino with plasma 
is valid~\cite{Ruffert:1997,Kuznetsov:1997,Birkl:2007}.

In Ref.~\cite{Beloborodov:2011}, the formula (10) for the energy deposition rate was taken, which 
was calculated in the crossed field limit~\cite{Kuznetsov:1997}. By the way, the value $q^\alpha$ 
defined by Eq.~(10) of Ref.~\cite{Beloborodov:2011} is not the 4-vector while the value 
$Q^\alpha = E \, q^\alpha$ is. 
However, in the region of the physical parameters 
used in Ref.~\cite{Beloborodov:2011} ($B$ to 180 $B_e$, $E$ to 25 MeV), the approximation of a crossed field is poorly applicable,
as well as the approximation of a superstrong field when $e^- e^+$ are created in the ground Landau level.
The contribution of the next Landau levels which can be also excited, should be taken into account.
We present here the results of our calculation of the mean neutrino energy losses caused by
the process $\nu \to \nu e^- e^+$ in a moderately strong magnetic field, i.e. in the conditions 
of the Kerr black hole accretion disk.  

We parametrize the energy deposition rate as:
\begin{equation}
Q_0 \, = \, (C_V^2 + C_A^2) \, \sigma_0 \, m_e^4\,E \; f \! \left( \frac{E}{m_e}\,, \frac{B}{B_e} \right) \,, 
\label{eq:Q0_2}
\end{equation}
where 
$\sigma_0 = 4 \, G_{\mathrm{F}}^2 \, m_e^2/\pi$, and the dependence on the initial neutrino energy and the field 
strength is described by the function $f ( {E}/{m_e}\,, {B}/{B_e} )$.
This function calculated in Ref.~\cite{Kuznetsov:1997} in the crossed field limit had the form
\begin{equation}
f^{(cr)} (y, \eta) \, = \, \frac{7\, y^2 \, \eta^2 }{1728 \, \pi^2} \, \ln (y \, \eta) \,, 
\label{eq:f_(cr)}
\end{equation}

On the other hand, in the strong field limit when both electron and positron are born in the ground Landau level, 
the function $f (y, \eta)$ was also calculated in Ref.~\cite{Kuznetsov:1997} and can be presented in the form
\begin{equation}
f^{(00)} (y, \eta) \, = \, \frac{\eta \, y^4}{32 \, \pi^2} \, 
\int\limits_0^1 \D \rho\, \rho (1 - \rho^2)^2 \,
\exp \! \left( - \frac{y^2 (1 + \rho^2)}{2 \eta} \right)
I_0 \! \left( \frac{y^2}{\eta} \, \rho \right),
\label{eq:f_(00)}
\end{equation}
where $I_0(x)$ is the modified Bessel function.

In conditions of moderately strong magnetic field, when the electron and the positron are created in the process
$\nu \to \nu e^-_{(n)} e^+_{(\ell)}$ in the $n$th and $\ell$th Landau levels, the result has more complicated form. 
It is significantly simplified when one of
the particles, electron or positron, is born in the ground Landau level, and if an approximation $B \gg B_e$ is used. 
We obtain the contribution of the channels $\nu \to \nu e^-_{(n)} e^+_{(0)}$ and $\nu \to \nu e^-_{(0)} e^+_{(n)}$ to the function $f (y, \eta)$ as follows:
\begin{eqnarray}
&&f^{(n0+0n)} (y, \eta) \, = \, \frac{\eta \, y^4}{4 \pi^2 (n-1)!} \, 
\left( \frac{y^2}{2 \eta} \right)^{n-1}
\int\limits_0^{1-\sqrt{b_n}} \D \rho\, \rho \, \int\limits_0^{Z_0} \frac{\D z (1-r)}{r (1-2 r + \rho^2)^2} 
\nonumber\\[2mm]
&& \times 
\left[(1 - \rho^2)^2 + 4 r^2-2 r (1+\rho^2) \right] 
\int\limits_0^{2 \pi} \frac{\D \phi}{2 \pi} (r-\rho \cos \phi) 
\nonumber\\[2mm]
&& \times \,
 (1 - 2 \rho \cos \phi + \rho^2)^{n-1} 
\exp \! \left( - \frac{y^2 (1 - 2 \rho \cos \phi + \rho^2)}{2 \eta} \right),
\label{eq:f_(0n)}
\end{eqnarray}
where 
\begin{eqnarray}
b_n = \frac{2 n e B}{E_{\mprp}^2}, \qquad
Z_0 = \frac{1}{2} \, \sqrt{\left(1 + \rho^2 - b_n \right)^2 - 4 \rho^2}\,.
\label{eq:notat}
\end{eqnarray}

\begin{figure}[ht] 
\centering 
\includegraphics*[width=0.87\textwidth]{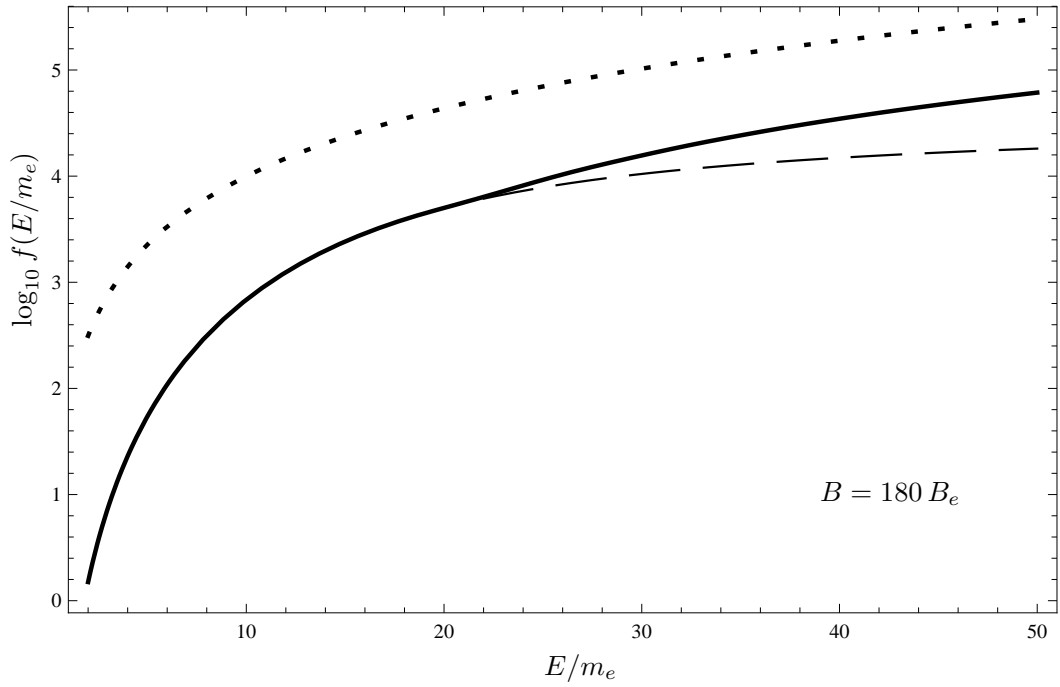}
\caption{The function $f (E/m_e)$ for $B=180 \,B_e$ obtained in the crossed field limit (dotted line), 
with $e^- e^+$ created at the ground (0,0) Landau level (dashed line), and for the sum of all lower 
Landau levels which are excited in this energy interval according 
to the condition~(\ref{eq:condE})(solid line).}
\label{fig:function180}
\end{figure}

\begin{figure}[!b] 
\centering 
\includegraphics*[width=0.87\textwidth]{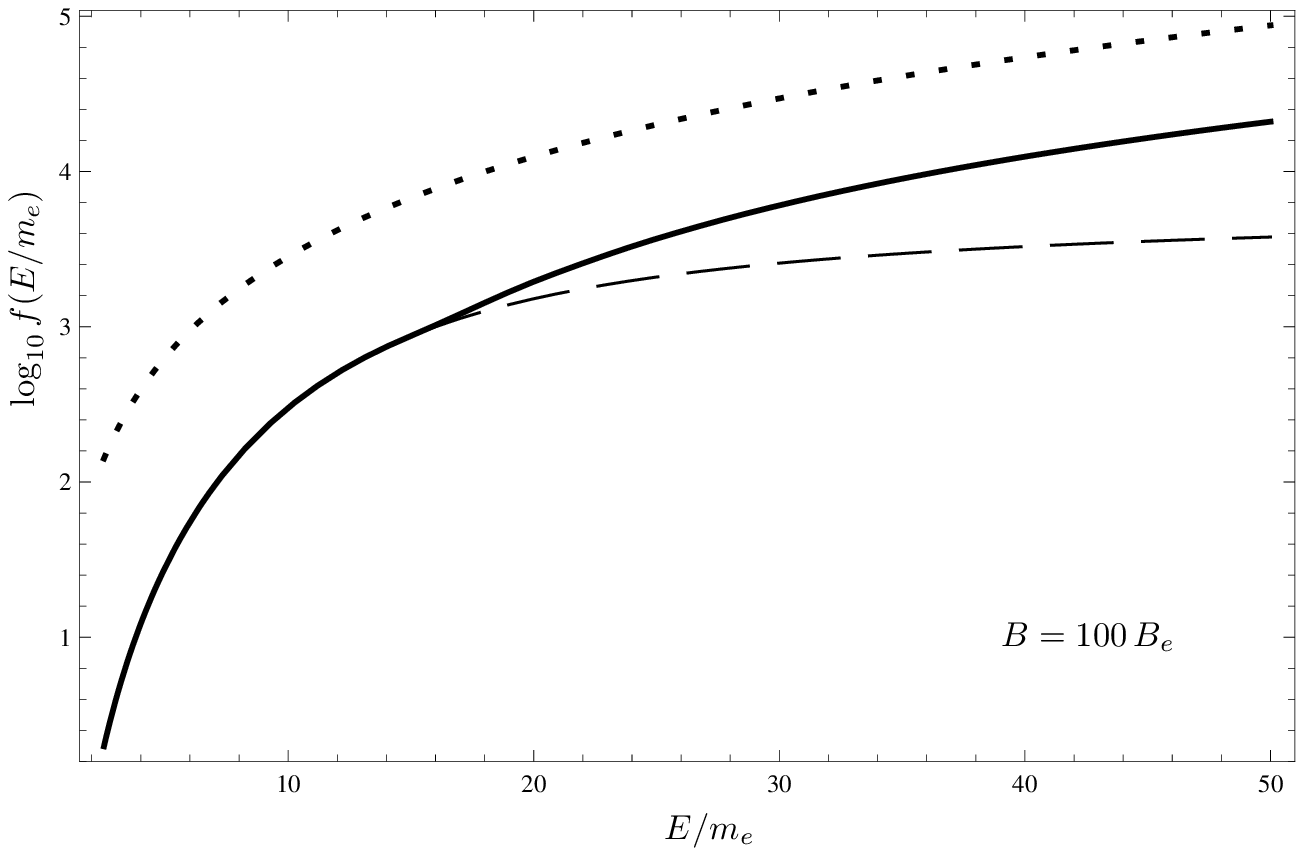}
\caption{The same as in Fig.~\ref{fig:function180}, for $B=100 \,B_e$.}
\label{fig:function100}
\end{figure}

\begin{figure}[ht] 
\centering 
\includegraphics*[width=0.87\textwidth]{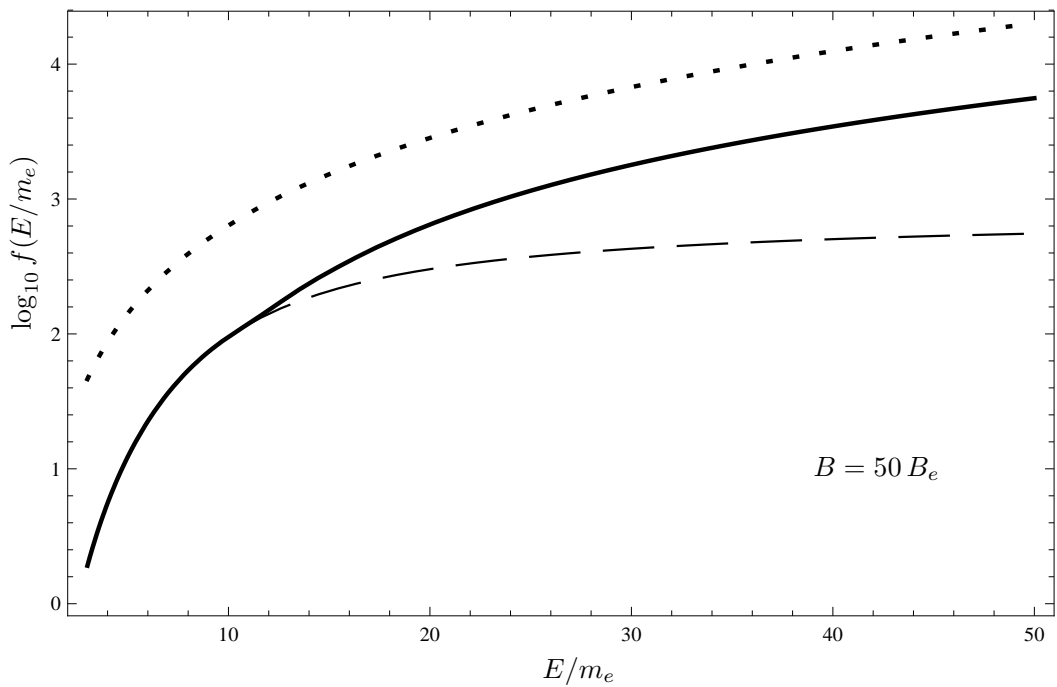}
\caption{The same as in Fig.~\ref{fig:function180}, for $B=50 \,B_e$.}
\label{fig:function50}
\end{figure}

In Figs.~\ref{fig:function180}--\ref{fig:function50}, the function $f (y, \eta)$ obtained in different approximations 
is shown at $B = 180 B_e, \,100 B_e, \,50 B_e$. 
It can be seen that the crossed field limit gives the overstated result which is in orders of magnitude greater 
than the sum of the contributions of lower excited Landau levels.   
On the other hand, the results with $e^- e^+$ created at the ground Landau level give the main contribution to 
the energy deposition rate, and almost exhaust it at $B = 180 B_e$. 

This would mean that the conclusion~\cite{Beloborodov:2011} that the contribution of the process 
$\nu \to \nu e^- e^+$ to the efficiency of the electron-positron plasma production by neutrino exceeds
the contribution of the annihilation channel $\nu \bar\nu \to e^- e^+$, and that the first process 
dominates the energy deposition rate, does not have a sufficient basis.  
A new analysis of the efficiency of energy deposition by neutrinos through both processes, $\nu \bar\nu \to e^- e^+$ 
and $\nu \to \nu e^- e^+$, in a hyper-accretion disc around a black hole should be performed, with taking 
account of our results for the process $\nu \to \nu e^- e^+$ presented here. 

\section{Conclusions}	
\label{sec:Conclusions}

In the paper, a calculation is performed of the mean value of the neutrino energy loss due to the process 
of electron-positron pair production, 
$\nu \to \nu e^- e^+$, in the magnetic field of an arbitrary strength at which the electrons and positrons 
can be produced in the states corresponding to the excited Landau levels, which could be essential
in astrophysical applications.  
The results obtained should be used for calculations of the efficiency of the electron-positron plasma production 
by neutrinos in the conditions of the Kerr black hole accretion disk, regarded by experts as the most likely source 
of a short cosmological gamma-ray burst. In those conditions, the crossed field limit used in the previous 
calculations led to the overstated result which was in orders of magnitude greater 
than the sum of the lower Landau levels. 
The study may be also useful for further development of computational techniques for the analysis 
of quantum processes in an external active environment, particularly in conditions of moderately strong magnetic field, 
when the allowance for the contribution of only the ground Landau level is insufficient.  

\section*{Acknowledgements}

We dedicate this paper to the blessed memory of our teacher, colleague, and friend 
Nickolay Vladimirovich Mikheev, who passed away on June 19, 2014. 

A.K. and D.R. express their deep gratitude to the organizers of the 
Seminar ``Quarks-2014'' for warm hospitality.  

\newpage

%

\newpage

The study was performed with the support by the Project No.~92 within the base part of the State Assignment 
for the Yaroslavl University Scientific Research, and was supported in part by the 
Russian Foundation for Basic Research (Project No. \mbox{14-02-00233-a}).



\end{document}